\documentclass[%
twocolumn,
superscriptaddress,
preprintnumbers,
 amsmath,amssymb,
prd,
]{revtex4-2}

\usepackage{graphicx,color}
\usepackage{dcolumn}
\usepackage{bm}
\usepackage{ulem}
\usepackage{color} 

\begin{document}


\title{Crossover equation of state based on color-molecular-dynamics }

\author{Nobutoshi Yasutake}
\email{nobutoshi.yasutake@p.chibakoudai.jp}
\affiliation{Department of Physics, Chiba Institute of Technology, 2-1-1 Shibazono, Narashino, Chiba 275-0023, Japan}
\affiliation{Advanced Science Research Center, Japan Atomic Energy Agency, Tokai, Ibaraki 319-1195, Japan}

\author{Toshiki Maruyama}
\email{maruyama.toshiki@jaea.go.jp}
\affiliation{Advanced Science Research Center, Japan Atomic Energy Agency, Tokai, Ibaraki 319-1195, Japan}



\date{\today}

\begin{abstract}
The equation of State for dense matter is studied with color molecular dynamics, in which hadron matter and quark matter are automatically distinguished only from quark color state. The quark-quark interactions are optimized to be consistent with saturation properties: symmetric energy, $L-$parameter, and incompressibility around nuclear density. In the calculations, the degrees of freedom of colors are solved at each numerical step, although the flavors are fixed as up or down quarks. The resultant mass-radius relations also satisfy the observational constraints such as the gravitational wave observations, NICER, and  ``{\it the two-solar mass observations}" of neutron stars. 
In this model with the allowed parameter range, deconfined quark matter appears in the core of neutron stars via crossover.
Although the current constraints from the observations are not enough to conclude whether quark matter appears at high-density region and also our study is still in a qualitative level, our method would help to understand high-density material properties inside neutron stars in the future. 
\end{abstract}


\maketitle

\section{\label{sec:level1}Introduction}
The equation of state~(EOS) is a key to understand neutron star~(NS) physics. In particular, it is an interesting and big question how hadron matter composed of baryons deconfine to quark matter at high density. However, since the lattice QCD simulations are not yet feasible for finite density and low-temperature $T\simeq 0$, there remains significant theoretical uncertainty regarding the behavior of matter inside NSs. It is a fundamental issue to clarify deconfinement behaviors. 

Instead of solely relying on theoretical and direct understanding of the confinement-decontainment phase changes, 
the investigation of astronomical observations and nuclear experiments has emerged as a promising approach.
In this context, it is the first step to reproduce the mass-radius~(MR) relations of NSs.
One of the well-known constraints on the MR relations is ``{\it two solar mass constraints}", which was derived from the observation of PSR J0348+432~\cite{Antoniadis2013}.
Gravitational wave observations also provide strong constraints on EOS. 
The observation GW170817 has yielded an upper limit of the radius of neutron stars assuming a binary system with canonical masses~\cite{Abbott2017}. The electromagnetic counterpart observations of the gravitational wave event have further constrained the MR relations of NSs~\cite{Bauswein2017}.
On the other hand, Neutron star Interior Composition Explorer~(NICER) has gradually narrowed the MR regions of NSs, 
e.g. PSR J0030+045~\cite{Riley2019,Miller2019} and PSR J0740+6620~\cite{Riley2021,Miller2021}.
Consequently, it is expected that astrophysical observations will significantly constrain the EOSs for high-density regions of NSs.

Thanks to these developments in astronomical observation techniques, it would be nice to consider the possibility of the quark-hadron phase change in NSs. This possibility has been pointed out by past studies, and many of them have assumed a first-order phase transition~\cite{Yasutake2009, Chen2013, Yasutake2014, Maslov2019, Xia2019, Xia2020}. The phase transitions, in this case, apply to that of multicomponent systems, where two phases are in equilibrium with different chemical compositions. In such phase transitions nonuniform structures, namely ``{\it pasta structures}" may appear, after the balance between the charge interactions and surface tensions. Not only the first-order phase transitions but also the possibility of the crossover are discussed~\cite{Masuda2016, Baym2019, Minamikawa2021, Blaschke2022,QHC21A}.

One of the points that we should keep in mind is that the model should describe nuclear matter at lower densities as a hadron phase and at higher densities as a quark phase, on the latter phase we do not have enough information. Furthermore, big uncertainties lie in the mechanisms of confinement and deconfinement, and the EOS in between. At lower densities the property of hadronic matter is rather well known: the binding energy per baryon of symmetric nuclear matter, $S_0$, has the minimum value of -16 MeV at the saturation density $n_0$, around 0.16 fm$^{-3}$. The range of symmetric energy $J$, which is the energy difference per baryon between the symmetric nuclear matter and the pure  neutron matter, is relatively narrowly restricted: a plausible range for $J$ of 29-33 MeV \cite{Tsang2012,Lattimer2013}. However, the values of the $L-$parameter (the slope of energy for neutron matter), and the incompressibility $K$ (second derivative of symmetric matter energy by the density) are strongly model dependent.

The fiducial range of the $L$-parameter is reported as  $L=60 \pm 20$ MeV, which is  constrained by analyses of terrestrial experiments, as summarized well in Refs.\cite{Vinas2014, Li2019}. 
Compared with them, there are also the possibilities for larger $L$-values: 
Danielewicz {\it et al.} have derived the range as $70 < L < 101$ MeV from the isobaric analog states and isovector skin results \cite{Danielewicz2017}. Radioactive Isotope Beam Factory (RIBF) at RIKEN in Japan has reported  as $42 < L < 117$ MeV \cite{Estee2021}. The Pb radius experiment (PREX)-II projects have suggested the range of $L$ as $L=106 \pm 37$ MeV \cite{Reed2021}, while the analyses after the Ca radius experiment (CREX) result have suggested the ordinal $L$-values recently. Hence, the current terrestrial experiments did not solely conclude the value of $L$.

It is well known that the $L-$parameter described above is strongly correlated with neutron skin thickness. It was theoretically predicted that the neutron skin thickness would change by considering alpha-clustering at the surface of neutron-rich heavy nuclei.
This point is mentioned in the paper on EOS by Typel {\it et al.} which is often used in numerical simulations in theoretical astronomy and widely known as ``DD2"~\cite{DD2}.
Recently, this prediction was confirmed experimentally by Tanaka {\it et al.}\cite{Tanaka2021}.
In this study, DD2 EOS is compared to our EOS as one benchmark.

As for the incompressibility $K$, Danielewicz {\it et al.}\citep{Danielewicz2002} pointed out 167 MeV $<K<$ 300 MeV, where they analyzed the flow of matter to extract pressures in nuclear collisions. Piekarewicz has given the range of $K= 248 \pm 8$ MeV from the iso-scalar giant monopole resonance (ISGMR) in $^{208}$Pb adopted the relativistic mean-field model with a random-phase-approximation~\cite{Piekarewicz2004}. G. Col\`{o} {\it et al.} have also analyzed the measurements of the ISGMR in medium-heavy nuclei introducing some types of Skyrme forces, then predicted the $K$ around 230 MeV~\cite{Colo2004}. As for the prediction of a small value of $K$,  Sturm {\it et al.} and Hartnack {\it et al.} have concluded $K$ is around 200 MeV from a comparison of the results of transport theories with the experimental data of heavy-ion collisions with the production of $K^+$ mesons \cite{Sturm2001,Hartnack2006}.

The EOS of the deconfinement quark phase should be consistent with all constraints as described above. 
For this purpose, Kojo {\it et al.} constructed EOSs describing quark deconfinement with crossover~\cite{QHC21A}. These EOSs are based on hadron EOSs derived from the chiral effective field theory nuclear EOS~\cite{Lonardoni2020, Drischler2021a, Drischler2021b} at low densities. By interpolating this EOS with the quark EOS at high density, they construct a crossover EOS. 
Such an interpolative method is practical as a first step to obtain EOS, but one should be careful to obtain thermodynamic quantities such as thermal conductivity since one cannot clearly distinguish hadron matter and quark matter in the crossover case. 
One possible approach to obtaining such physical quantities is to discuss the deconfinement-confinement mechanism only within the framework of the quark model.

Keeping this in mind, we employ a color molecular dynamics~(CMD) simulation which deals with constituent quarks with color degrees of freedom \cite{MaruyamaHatsuda2000}. In this method, we solve the time evolutions of positions, momenta, and color coordinates of quarks, which are governed by the Hamiltonian with the potential term consisting of the color confining potential, the perturbative gluon-exchange potential, and the meson-exchange potentials. 

At low density, quarks are clusterized into baryons, in which the color-singlet state is favored. As density increases, the baryons start to overlap with each other, and then they start deconfined into quark matter. Our CMD simulations show such percolative behavior during the deconfinement. Compared with our previous studies~\cite{Akimura2005}, we improved the scheme to include relativistic kinetic energy and solve the time-dependence of the color~\cite{MaruyamaHatsuda2000}. 
We also include color-independent nonlinear quark-quark repulsions to keep the consistency with mass-radius relations of compact stars constrained by the astrophysical observations. 
The nonlinear interaction can be understood as the quark many-body effects. 

This paper is organized as follows. 
In Section~II, we outline our framework for CMD. 
Section~III contains numerical results consistent with the constraints of mass-radius relations from astronomical observations, and with saturation properties around nuclear density. 
Section~IV is devoted to the conclusion and the discussion of our results.

\section{\label{sec:level2}Color Molecular Dynamics}

Our formulation is based on our previous papers \cite{MaruyamaHatsuda2000, Akimura2005}. 
Throughout this paper, strangeness is not considered, and it is assumed that both $u$- and $d$-quarks have the constituent quark mass, $m$.

We start with the total wave function defined by a direct product of single-particle wave packets of quarks, the position and the momentum of which are centered at time-dependent parameters ${\bf R}_i$ and ${\bf P}_i$, respectively, and $\chi_i$ is the internal degree of freedom given by a direct production of the fixed flavor ${\chi_i}^f$, the time-dependent color ${\chi_i}^c$, and the fixed spin orientation ${\chi_i}^s$,
\begin{equation}
\begin{split}
\Psi &( {\bf r}_1, {\bf r}_2, ...{\bf r}_N )=\\
&\prod_{i=1}^{N} \frac{1}{(\pi {L_q}^2)^{3/4}}
\exp \left[ -\frac{({\bf r}_i-{\bf R}_i)^2}{2{L_q}^2}
+\frac{i}{\hbar}{\bf P}_i{\bf r}_i
\right]\chi_i,
\end{split}
\label{eq:Gaussian}
\end{equation}
where $N$ and $L_q$ denote the total number of quarks and the fixed width of wave packets, respectively. We employ the width $L_q=0.37$ fm in this work. 
Here, $r_i$ represents the coordinates of $i$th particle somewhere in the Gaussian wave packet centered at $R_i$. It is crucial not to confuse $r_i$ with $R_i$. In practice, $r_i$ does not explicitly appear during the calculations due to the double-folding integrals.
In this paper, we fix the flavors and spins, hence ${\chi_i}^f$ and ${\chi_i}^s$ are treated as constants. 
The explicit form of the time-dependent degree of freedom for color is shown as 
\begin{eqnarray}
{\chi_i}^c  = \left(
\begin{array}{c}
\cos\alpha_i \;e^{-i\beta_i}\;\cos\theta_i  \\
\sin\alpha_i \;e^{+i\beta_i}\;\cos\theta_i  \\
\sin\theta_i\;e^{i\varphi_i} \\
\end{array}
\right),
\label{eq:color}
\end{eqnarray}
hence $\alpha_i$, $\beta_i$, $\theta_i$, $\varphi_i$ are the variables for the color of each particle.

The system follows the Hamiltonian,
\begin{equation}
H = H_0+ V_{\rm Pauli}-T_{\rm spur},
\label{eq: H}
\end{equation}
where $H_0$ is the conventional Hamiltonian expressed as
\begin{equation}
H_0=\left<\Psi\left|
\sum_{i=1}^{N}\sqrt{m+\hat{\textbf{p}}^2_i}+ \hat  V_{\rm C} +  \hat  V_{\rm M}
\right| \Psi \right>. 
\end{equation}
The first term is the kinetic term with relativistic kinematics.
The second term is the interaction term expressed as
\begin{eqnarray}
\hat V_{C}&=&-\frac{1}{2}\sum_{i.j\ne i}^{N}\sum_{a=1}^8\frac{\lambda_i^a\lambda_j^a}{4}\left( \kappa \hat{r}_{ij}-  \frac{\alpha_s}{\hat{r}_{ij}} \right),
\label{eq: Vcol}\\
\hat V_{\rm M}&=& \frac{1}{2}\sum_{i}^{N} \left[
\frac{g_{\omega}^2C_\omega}{4 \pi}  \left(  \sum^N_{j \ne i} \frac{e^ {- \mu _\omega \hat{r}_{ij}}}{\hat{r}_{ij}} \right)^{1+\epsilon_\omega} \right. \nonumber \\
 &-&\left. \frac{g_{\sigma}^2C_\sigma}{4 \pi}  \left(  \sum^N_{j \ne i} \frac{e^ {- \mu _\sigma \hat{r}_{ij}}}{\hat{r}_{ij}} \right)^{1+\epsilon_\sigma} \hspace{-5mm}
 +\sum^N_{j \ne i}
 \frac{\sigma_i^3 \sigma_j^3}{4}\frac{g_{\rho}^2}{4 \pi}\frac{e^ {- \mu _{ \rho } \hat{r}_{ij}}}{\hat{r}_{ij}} 
\right] \nonumber\\
~~~
\label{eq: Vmeson}
\end{eqnarray}
where $\hat{r}_{ij}\equiv |\hat{\bf r}_i-\hat{\bf r}_j|$ is the distance between $i$th and $j$th quarks, 
and $\lambda^a_i$ is the Gell-Mann matrix.
The summation of $a$ in Eq.(\ref{eq: Vcol}) might be somewhat obscure, but considering the energy expectation value between two particles makes its significance more apparent: The potential between $i$, and $j$th particles is proportional to 
$\sum_{a=1}^8 \langle \chi^c_i| \lambda^a |\chi^c_i\rangle  \langle\chi^c_j|\lambda^a|\chi^c_j \rangle$.
The color-dependent interaction $\hat{V_C}$ consists of the linear confining potential (the first term) and the one-gluon exchange potential (the second term) as shown in Eq.(\ref{eq: Vcol}). The string tension of confinement $\kappa$ and the QCD fine structure constant $\alpha_s$ are set as $\kappa=0.75$ GeV~fm$^{-1}$, and $\alpha_s=1.25$, as shown in Ref.~\citep{MaruyamaHatsuda2000}.
Note that these two interactions are the main components to determine the nucleon mass $M$, and these parameter sets are typical values \citep{Yoshimoto2000}. In this method, we do not conduct the anti-symmetrization of the total wave function, in which the numerical cost is $N^4$ \cite{Feldmeier1990, 
Ono2005}. Hence, the interaction is underestimated by a factor of 4 when we take the matrix element of $\tau^a_i \tau^a_j$, and then we introduce the effective coupling constants as $\kappa_{\rm eff}=4 \kappa $ and $\alpha_{s}^{\rm eff}=4 \alpha_s$ to get consistency with color SU(3) algebra. The constituent quark mass is then obtained as $m=361.8$ MeV, which is determined to reproduce the nucleon mass, $M=938$ MeV, with the above parameter sets for the confinement and the one-gluon exchange.

As for the nuclear force, the nonperturbative gluon exchanges in the color singlet channels would be the essential part.
Hence, we introduce the $\sigma+\omega+\rho$ quark-meson couplings acting between quarks~\cite{Guichon1988}.
These coupling constants are set as $g_{\omega}=5.46$, $g_{\sigma}=3.23$, and $g_{\rho}=8.19$.
The meson-quark coupling constants $g_{\omega}$, $g_{\sigma}$, and $g_{\rho}$, are estimated from the meson-nucleon couplings as $g_{\omega}=g_{\omega N}/3=4.98$, $g_{\sigma}=g_{\sigma N}/3=3.09$, and $g_{\rho}=g_{\rho N}/3 =9$, in previous works~\cite{Akimura2005}. These values of $g_{\omega}$, $g_{\sigma}$, and $g_{\rho}$ in this paper are different by $\sim$10 \% from our previous works. We also introduce the small nonlinearity parameters $\epsilon_\omega$, and $\epsilon_\sigma$ for the $\omega$-, and $\sigma$-exchange potentials.
The physical significance of this nonlinearity is the effects of many-body correlations beyond two-body interactions. When $\epsilon_\omega$ and $\epsilon_\sigma$ are zero, it exactly coincides with two-body correlations for $i$- and $j$-particles. However, when they are not zero, we have to take into account the contribution from all particles to the correlation between  $i$- and $j$-particles. This expression might raise concerns about an exceptionally increased computational cost. However, that is not the case. The summation part of this equation can be precalculated outside the loop operation that calculates the two-particle correlation energy. By doing so, calling this sum again within the loop for two-particle correlation results in a computational cost of $N^2$. In this manner, it is possible to incorporate many-body effects while keeping the numerical cost low.
We also introduce $C_{\omega}$, $C_{\sigma}$to make the coupling constant $g_{\omega}$, $g_{\sigma}$ dimensionless, set as $C_{\omega}=1/(1+{\epsilon_\omega})$, $C_{\sigma}=1/(1+{\epsilon_\sigma})$, which are same as Ref.\cite{Akimura2005}. We have chosen $\epsilon_\omega=0.20$, $\epsilon_\sigma=-0.13$ to make our EOS consistent with the constraints from astrophysical observations and experimental nuclear physics described later. 
We have introduced the effective widths $L_\omega$, $L_\sigma$, and $L_\rho$ as $L_\omega=$ 0.75 fm, $L_\sigma=1.35$ fm, and $L_\rho=$ 1.30 fm for each meson exchange term. 
Also, these values are almost the same as the previous work: $L_\omega=$ 0.70 fm, and $L_\sigma=L_\rho=$ 1.30 fm~\cite{Akimura2005}.
The values $L_\omega$, $L_\sigma$, and $L_\rho$ are set to larger values than $L_q$ so that the quark-meson interactions replicate the nucleon-meson interactions.

Here, we discuss the effective ranges of each interaction. As indicated in Eq.(\ref{eq:Gaussian}), we assume a Gaussian wave packet for each particle. The convolutions of the potential energies are then calculated through the double-folding integrations. 
Thus, the total energy is determined by these interactions, except for the confinement potential, within an effective range dependent on $L_\omega$, $L_\sigma$, $L_\rho$, and $L_q$. 
Therefore, by choosing a sufficiently large periodic-boundary, the correlation energies with long distances become negligibly small, though we compute the potentials for all correlations without approximation.
In this study, we ensure that all particles form a color singlet. Consequently, the potentials related to color, namely $\hat{V_C}$, including the confinement potential, are kept finite in total. However, a box-size dependency may still exist. Therefore, it would be necessary to investigate the box size dependence systematically. But due to computational costs, we will not go into detail and describe them in the appendix.

Instead of the anti-symmetrization of the wave function, we introduce an effective Pauli-potential, $V_{\rm Pauli}$. It acts as a repulsive force between quarks with the same intrinsic degree of freedom such as flavor, color, and spin, collectively denoted as $\chi_i$. 
According to our previous studies, we employ the following form of Pauli potential, 
\begin{eqnarray}
V_{\rm Pauli}=   \frac{1}{2}\sum_{i,j \ne i}^{N} &&
\frac{C_p}{(q_0 p_0)^3}\exp{\left[ -\frac{({\bf R}_i-{\bf R}_j)^2}{2q_0^2}\right]} \nonumber\\
&&\times \exp{\left[ -\frac{({\bf P}_i-{\bf P}_j)^2}{2p_0^2}\right]} \delta_{\chi_i,\chi_j} 
\label{eq: Pauli}
\end{eqnarray}
The parameters are set as $q_0 = 2.46$ fm, $p_0 = 240$ MeV, and $C_p =$ 131 MeV, to reproduce relativistic kinetic energy for free fermions at zero temperature. Pauli potential is highly phenomenological, so that we did not write it in terms of $r_i$ but by the central values of position and momentum, $R_i$ and $P_i$.

In this paper, we do not take into account the spurious zero-point energy, $T_{\rm spur}$, which comes from the center-of-mass motion of clusters as shown in Ref.\cite{Akimura2005}. These effects remain future problems.
The time evolution of the system is given by the Euler-Lagrange equation for \{${\bf R}_i$, ${\bf P}_i$, $\alpha_i$, $\beta_i$, $\theta_i$, 
$\varphi_i$\} with the classical Lagrangian \cite{MaruyamaHatsuda2000}. Accordingly,
the explicit form of the equations of motion reads:
\begin{eqnarray}
\dot {\bf R}_i
&=&{\partial H\over\partial {\bf P}_i}, \ \ \ \ 
\dot {\bf P}_i
=-{\partial H\over\partial {\bf R}_i}, \label{eq: rp_evo}\\
\dot\beta_i
&=&-{1\over2\hbar\sin2\alpha_i\;\cos^2\theta_i}{\partial H\over\partial\alpha_i},\\
\dot\theta_i
&=&{1\over \hbar\sin{2\theta_i}}{\partial H\over\partial\varphi_i},\\
\dot\alpha_i
&=&{1\over2\hbar\sin2\alpha_i\;\cos^2\theta_i}{\partial H\over\partial\beta_i}
   -{\cos2\alpha_i\over2\hbar\sin2\alpha_i\;\cos^2\theta_i}{\partial H\over\partial\varphi_i}, \nonumber \\
~~~\\
\dot\varphi_i
&=&-{1\over \hbar\sin{2\theta_i} }{\partial H\over\partial\theta_i}
  +{\cos2\alpha_i\over2\hbar\sin2\alpha_i\;\cos^2\theta_i}{\partial H\over\partial\alpha_i}.
\end{eqnarray}

In the calculations, all quarks are, initially, distributed randomly without momenta in a box with the periodic boundary condition.
The ground state (matter at zero temperature) is obtained by the frictional cooling~\cite{MaruyamaHatsuda2000, Akimura2005}.
For this purpose, we solve a damping equation of motion instead of Eq.(\ref{eq: rp_evo}),
\begin{eqnarray}
& & \dot {\bf R}_i={\partial H\over \partial {\bf P}_{i}}  +\mu_R {\partial H\over \partial {\bf R}_{i}} \label{eq: rfric}, \\
& & \dot {\bf P}_i=-{\partial H\over \partial {\bf R}_{i}}  +\mu_P {\partial H\over \partial {\bf P}_{i}} \label{eq: pfric}, 
\end{eqnarray}
 where $\mu_R$ and $\mu_P$ are damping coefficients
 set as $\mu_R=-0.00002$ and $\mu_P=-0.02$.
The values of $\mu_R$ and $\mu_P$ can, in principle, be arbitrary; however, reducing them comes at the cost of increased numerical complexity. On the other hand, smaller values make it easier to avoid falling into a local energy minimum state. The values adopted in this study are the optimal settings that show numerical convergence efficiently. In other words, for smaller values of $\mu_R$ and $\mu_P$, the results would still converge to the similar energy values in the end.

Based on the above formulation, we search for parameter sets consistent with the constraints from astronomical observations and nuclear experiments. For this purpose, we performed iterative calculations under fixed periodic boundary conditions with a size of 6 fm. Although it would be possible to investigate size dependence by changing the size or preparing a larger cell size, we refrained from doing so in this study due to the significant increase in computational cost. For instance, doubling the computational domain size requires an 8 times increase in the number of particles for the same density, resulting in an $8^2$ times increase of correlations. Even if such calculations were repeated to obtain one data point for the EOS at a certain density, it would be inefficient ignoring some physics to be considered, such as the color-magnetic interactions.

Once the parameters are given, the energy per nucleon for symmetric nuclear matter and neutron matter are obtained. 
Then, we have made fits of the ground-state energies by regressions.
Using the regressions, we calculate the EOS of matter in the charge neutral and beta-equilibrium including the contribution of electrons. 
Thanks to GPU parallel computing, it takes just one day to check one parameter set. 
In molecular dynamics, the heaviest numerical cost is in the particle correlation part.
There are some well-known techniques to speed up calculations in molecular dynamics, such as the tree method and fast multipole method (FMM) \cite{Cheng1999}, but, to our knowledge, there does not exist such methods involving confining potential, which does not decrease at a long distance, hence we do not use such coarse-graining techniques.

\section{RESULTS}

\subsection{Equation of state by color molecular dynamics}

\begin{figure}[tb]
\includegraphics[width=0.47\textwidth]{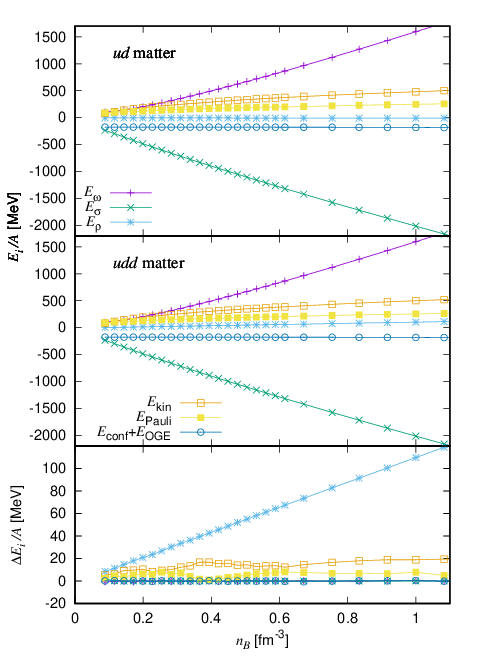}
\caption{\label{fig: Ei} Energy components per nucleon for $ud$ matter (upper panel) and $udd$ matter (middle panel), and the difference between them (lower panel).
Each dot represents the numerical results by CMD. 
The energies, which come from quark-meson couplings of $\omega$, $\sigma$, $\rho$, are shown as $E_\omega$, $E_\sigma$, $E_\rho$. 
The captions, $E_{\rm kin}$ and $E_{\rm Pauli}$, are the quark kinetic energy and the quark Pauli energy. The energies originated from the confinement and the one-gluon exchange are expressed as $E_{\rm conf}$ and $E_{\rm OGE}$. }
\end{figure}

Let us first show the energy components per nucleon for $ud$ and  $udd$ matter, and the difference between them in Fig.\ref{fig: Ei}. In symmetric nuclear matter, $u$ and $d $quarks are equally present, hence we call it $ud$ matter. In the same way, neutron matter is called $udd$ matter in this paper.
The total energy consists of the potential energies of quark-meson coupling such as $q$-$\omega$, $q$-$\sigma$, and $q$-$\rho$, the confining potential, the one-gluon exchange potential, the Pauli potential, and the kinetic energy, expressed as $E_\omega$, $E_\sigma$, $E_\rho$, $E_{\rm conf}$, $E_{\rm OGE}$, $E_{\rm Pauli}$, and $E_{\rm kin}$, respectively.

At the low density known as the inner crusts of NSs, pasta structures appear in the approximate scale from 10 fm to 30 fm, due to the balance of the Coulomb repulsion and the surface tension. Hence, one should incorporate Coulomb interactions and  search for the optimal size by varying the periodic-box size, or perform calculations in a sufficiently large domain~\cite{Caplan2018}. As mentioned in the discussion on the confinement potential, such computations have not been conducted in the present study.

As the density increases, the main component of the energy changes to $q$-$\omega$ and $q$-$\sigma$ couplings which are nonlinearly dependent on the density, while color-dependent potentials are rather moderate since they have roughly linear dependence on the density.
Furthermore, the primary component distinguishing the energy between $ud$ matter and $udd$ matter is attributed to the $\rho$ meson, $E_\rho$.

\begin{figure}[tb]
\includegraphics[width=0.47\textwidth]{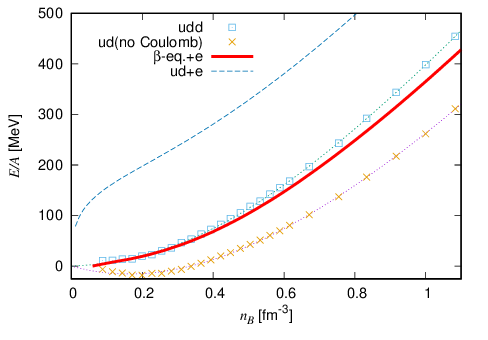}
\caption{\label{fig: eos} Density dependence of energy per baryon $E/A$ for each matter 
$ud$ matter and $udd$ matter.
Each dot represents the numerical results by CMD. Thin dot lines are the regression curves of them. The thick line, which is labeled by ``$\beta$-eq.+e", is obtained $E/A$ under the charge-neutral and beta-equilibrium conditions including contribution of electrons.
The dashed curve represents $E/A$ for $ud$ matter with electrons under the charge-neutral condition.
}
\end{figure}

The total energies of matter are shown in Fig.\ref{fig: eos}. Note that these quantities are obtained based on the quark models, namely CMD only.
In this figure, the regression curves are obtained assuming a sigmoid function with regularization terms as 
\begin{eqnarray}
(E/A)_{ud}(x)= 3\Big( a_1x + a_2x^2 ~~~~~~~~~~~~~~~~~~~~~~~~~\nonumber \\
+ \frac{a_3}{1 +  \exp(-a_4 x+a_5)} - \frac{a_3}{1+\exp(a_5)} \Big) , 
\label{eq: ud}
\end{eqnarray}
\begin{eqnarray}
(E/A)_{udd}(x)= 3\Big( b_1x^2+ b_2x^4 ~~~~~~~~~~~~~~~~~~~~~~~~~\nonumber \\
+  \frac{b_3}{1 +  \exp(-b_4 x+b_5)} - \frac{b_3}{1+\exp(b_5)}\Big), 
\label{eq: udd}
\end{eqnarray}
where $x$ is normalized baryon density as $x = n/{\bar n}$. 
Note that $\bar{n}$ is not necessarily the saturation density, but just a normalization factor set as ${\bar n}=0.16$ fm$^{-3}$.
Factor 3 at the beginning represents ``three quarks", i.e. per nucleon. In other words, by removing this factor, we can obtain the energy per quark, $E/Q$. The first two terms in both equations are introduced for the regularization of the regression.
The last terms are introduced to make the regression curves on the origin.
%
The obtained parameters are shown in Table \ref{tab: fitting}. 
The resultant fitting curves of energy per baron for $ud$ matter and $udd$ matter are also shown in Fig.\ref{fig: eos}.
From these curves, we can obtain the characteristic values for EOS around saturation density $n_0$: the saturation energy $S_0$, the symmetry  energy $J$, the density gradient of neutron-matter energy $L$, and the curvature of symmetric-matter energy $K$. 
They are summarized in Table \ref{tab: constraints}.

On the other hand, when we focus on high-density matter inside NSs, the charge-neutral condition and the beta-equilibrium are considered to be  realized. However, even for hadron matter, it is described by quark dynamics in this model. 

Hence, the charge-neutral condition is evaluated as, 
\begin{eqnarray}
0 = \sum_{i=u,d,e} Q_i n_i,
\label{eq: chneutral}
\end{eqnarray}
where $Q_i$, $n_i$ are the particle charge and density. We do not take into account muons for simplicity. 
In this study, we also neglect antiparticles, since we obtain ground state at zero temperature conducting the frictional coolings shown in Eq.(\ref{eq: rfric}) and Eq.(\ref{eq: pfric}). In Fig.\ref{fig: eos}, we show both cases of $ud$ matter with/without the contribution of Coulomb energy. The former one is obtained under the charge-neutral condition, and shown as the dashed curve in Fig.\ref{fig: eos}. As for the latter one, the role of electrons is necessarily important, hence it does not include electrons.

In addition, the beta-equilibrium must be also fulfilled between $u,d$ quarks and electrons in any phases:
\begin{eqnarray}
\mu_u+\mu_e = \mu_d,
\label{e: chemeq}
\end{eqnarray}
where $\mu_u$, $\mu_d$, and $\mu_e$ are the chemical potentials for $u$, $d$ quarks, and electrons, respectively. 

Once we obtain $(E/A)_{ud}$ and $(E/A)_{udd}$ curves, the energy per nucleon $E/A$ under the charge-neutral condition and beta-equilibrium is calculated as the mixture of $ud$ matter and $udd$ matter. 
\begin{eqnarray}
E/A &=& (1-\beta^2) (E/A)_{ud}+\beta^2 (E/A)_{udd}, 
\label{eq: energy}
\end{eqnarray}
where $\beta$ is defined to denote the asymmetry of $u,d$ quark density as
\begin{eqnarray}
\beta &=& 3\cfrac{n_d - n_u}{n_d + n_u}. 
\label{eq: beta}
\end{eqnarray}
In our calculation, we have to find the optimal value of $\beta$ consistent with both conditions at each density.
This result is also shown as the thick line in Fig.\ref{fig: eos}, including the energy contribution from electrons.

%
\begin{center}
\begin{table}[tb]
\begin{tabular}{|ccccc|}
\hline
 $a_1$ & $a_2$ & $a_3$  & $a_4$ & $a_5$ \\
 $b_1$ & $b_2$ & $b_3$ & $b_4$ & $b_5$ \\
 $\rm {[MeV]}$ & [MeV] & [MeV]&  & \\
\hline \hline
 -92.13  & 4.52325 & 1636.17 &  0.202968 &  0.263397  \\
 2.70704 & -0.00203469  & 38.0054  & 0.728765  & 2.83639  \\
\hline
\end{tabular}
\caption{The optimized parameters for Eq.(\ref{eq: ud}) and Eq.(\ref{eq: udd}).}
\label{tab: fitting}
\end{table}
\end{center}

%
\begin{center}
\begin{table}[tb]
  \begin{tabular}{|ccccc|}
     \hline
     $n_0$ & $J$ & $L$ & $K$ & $S_0$  \\
    $[\rm fm^{-3}]$ & [MeV] & [MeV]& [MeV] & [MeV]   \\
    \hline \hline 
    0.167 &31.0  & 74.2 & 260 & -15.8  \\
    \hline
  \end{tabular}
   \caption{The characteristic physical values around the saturation density obtained by CMD calculations. }
  \label{tab: constraints}
\end{table}
\end{center}

\begin{figure}[tb]
\includegraphics[width=0.49\textwidth]{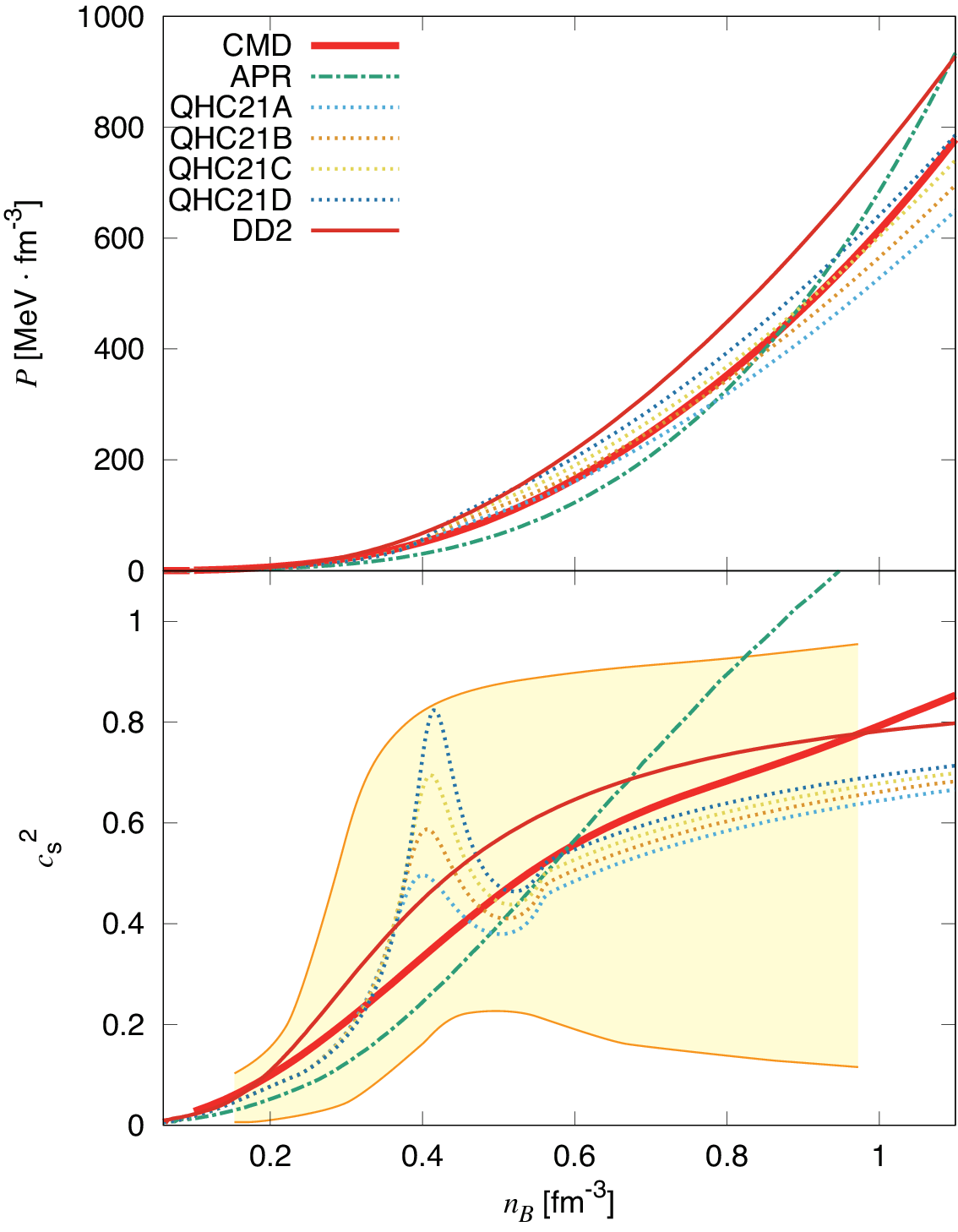}
\caption{\label{fig: p} 
The upper panel shows the pressure as a function of baryon density under charge-neutral and beta-equilibrium conditions, along with the results from Akmal {\it et al.}(APR)~\cite{APR}, Kojo {\it et al.}(QHC21A-D)~\cite{QHC21A}, and S. Typel {\it et al.}(DD2)~\cite{DD2} for comparison. 
The lower panel shows the corresponding sound speeds normalized by light speed.
The band (painted region) shows the constraint from ``PSR+GW +J0030+J0740". See the text for the details.
}
\end{figure}

In the upper panel of Fig.\ref{fig: p}, we show the relationship between pressure and baryon density corresponding $E/A$ 
under charge-neutral and beta-equilibrium conditions shown in Eq.(\ref{eq: energy}). 
However, below the subnuclear density $n_B < 0.06$ fm$^{-3}$, we use the EOS by Baym {\it et al.}(BPS)~\cite{Baym1971}. 
There is no density jump which is a characteristic of the first-order phase transition. This suggests that the deconfinement of quarks appears as the crossover. 
The EOSs by Akmal et al.(APR)~\cite{APR}, Kojo {\it et al.}(QHC21A-D)~\cite{QHC21A} and S. Typel {\it et al.}(DD2)~\cite{DD2}, available on the CompOSE archive~\cite{Comp}, are included in the figure for comparison.
The EOSs by Akmal {\it et al.}(APR) and S. Typel {\it et al.}(DD2) have often been adopted as benchmarks, but the EOSs by Kojo {\it et al.} are also chosen for comparison in this paper since they are the state-of-the-art crossover EOSs. 

The corresponding sound speeds normalized by light speed are shown in the lower panel of Fig.\ref{fig: p}.
The band (painted region) shows the constraint from ``PSR+GW +J0030+J0740" traced Ref.~\citep{QHC21A}, which has been originally deduced by Legred {\it et al.}\cite{Legred2021}. Since the EOSs except for APR are designed to be within this range, it is obvious that they satisfy the constraint. It can be seen that the sound velocity of our result is monotonically increasing with density, but not for QHC A-D models.
This difference is due to the quark deconfinement occurring over a wide density range in our crossover EOS, compared with QHC A-D models. 
Our results suggest that the pure quark matter does not appear even in the core of neutron stars with the maximum mass.
However, we confirm that the peak of sound velocity appears in our CMD calculations at high-density region, which is over the maximum density inside of stable NSs: $n_B = 1.08$ fm$^{-3}$.
Additionally, our model does not violate the causality law in the stable interior of neutron stars with respect to the speed of sound, but it does at higher density, $n_B=$ 1.42~fm$^{-3}$. When we extend our code to be fully relativistic in the future, it might be possible that it may no longer violate the causality law at such high densities.

\begin{figure*}[tb]
\includegraphics[width=0.23\textwidth]{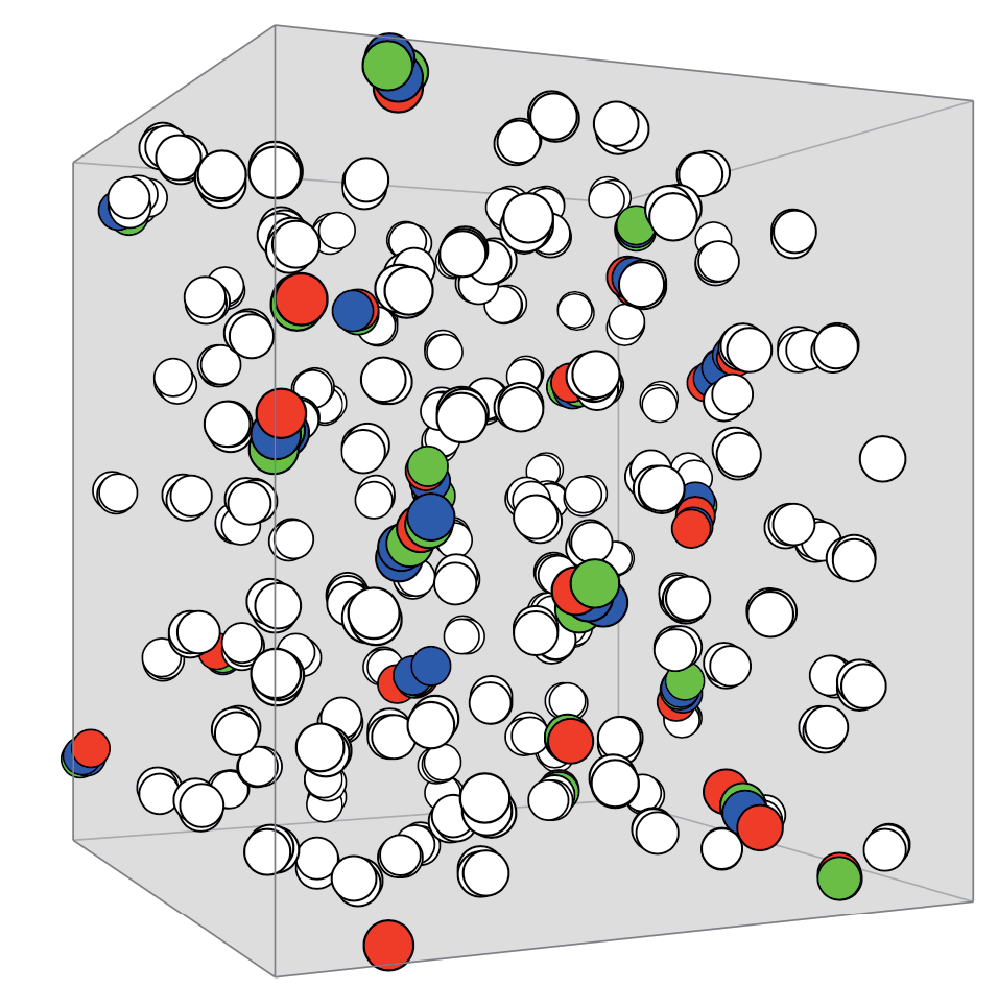}
\includegraphics[width=0.22\textwidth]{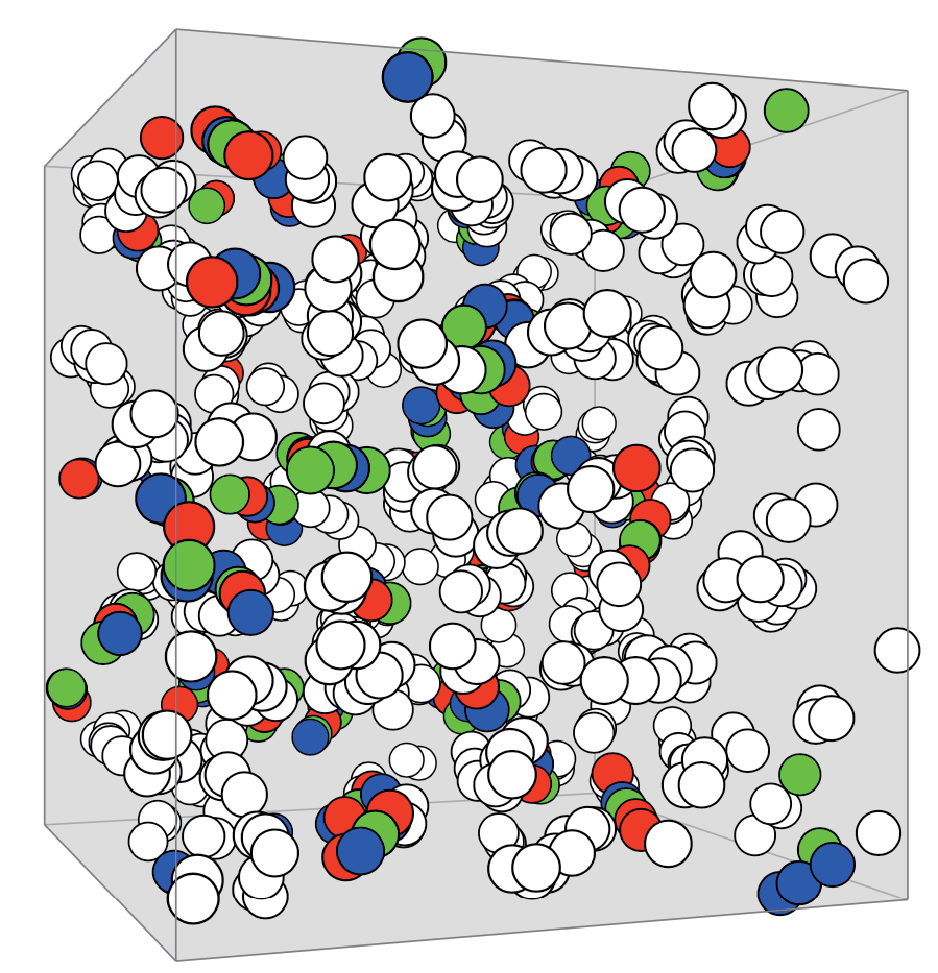}
\includegraphics[width=0.22\textwidth]{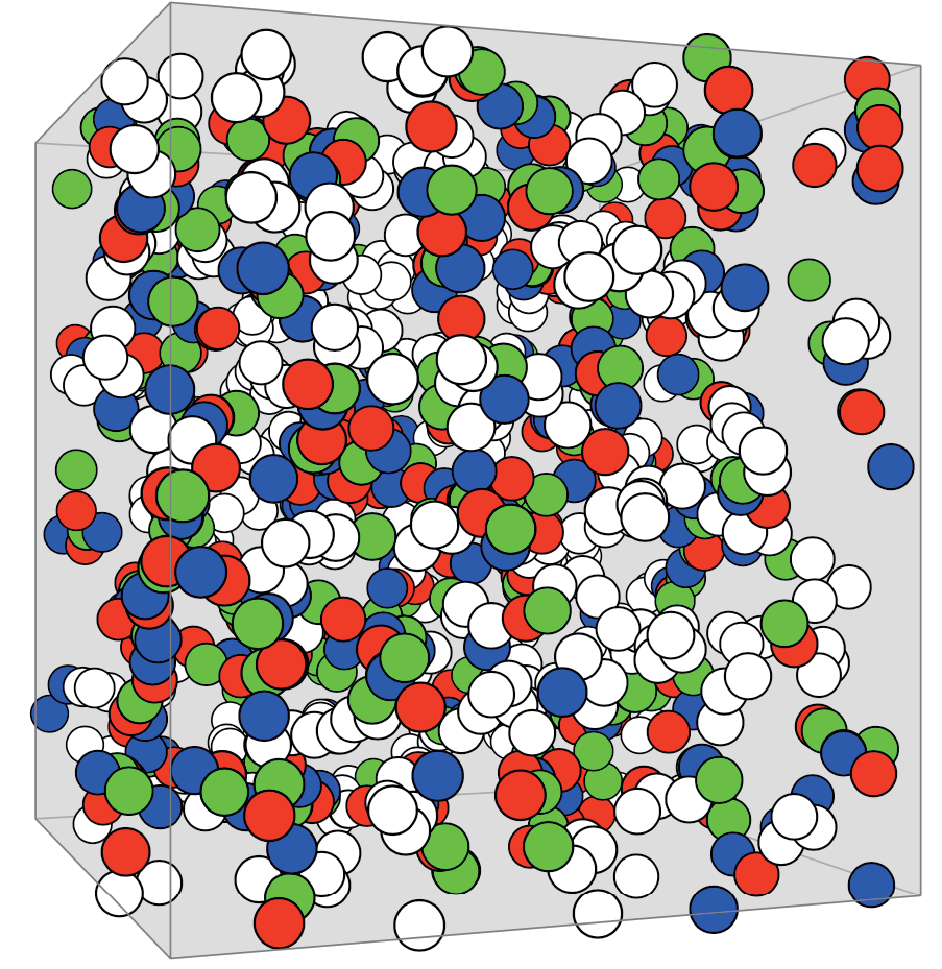}
\includegraphics[width=0.22\textwidth]{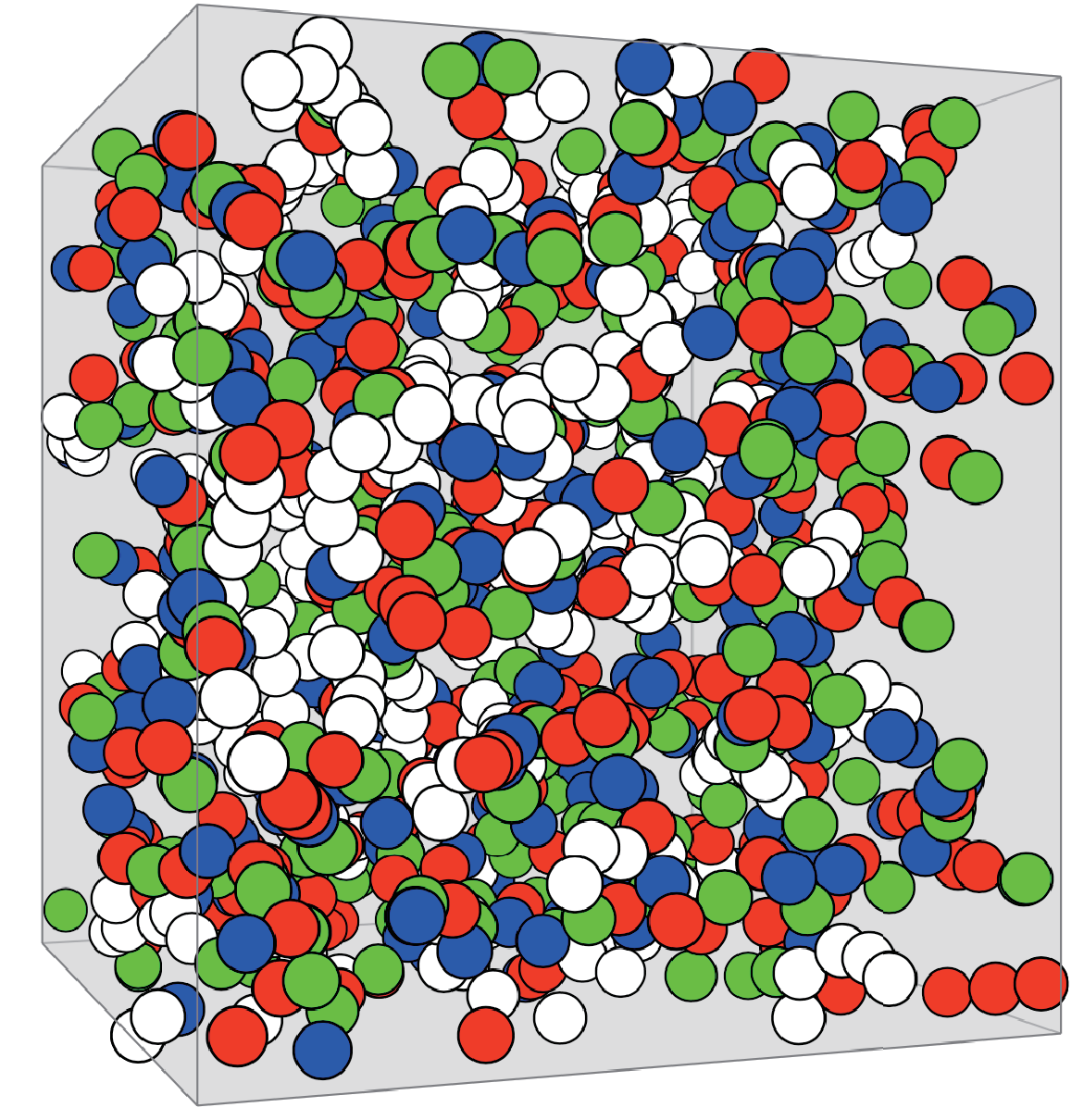}
\par
\vspace{-2.5mm}
\hspace{27mm} 0.833 fm$^{-3}$ \hspace{27mm} 1.166 fm$^{-3}$\hspace{24mm} 1.583 fm$^{-3}$ \hspace{24mm} 1.916 fm$^{-3}$
\caption{\label{fig:ud_cmd} The perspectives for $ud$ matter depended on density. Each color corresponds to the color's internal degree of freedom for each quark: the white color balls represent the quarks in the baryon state, while red, blue, and green ones do in the deconfined quark matter.}
\end{figure*}

\begin{figure*}[tb]
\includegraphics[width=0.22\textwidth]{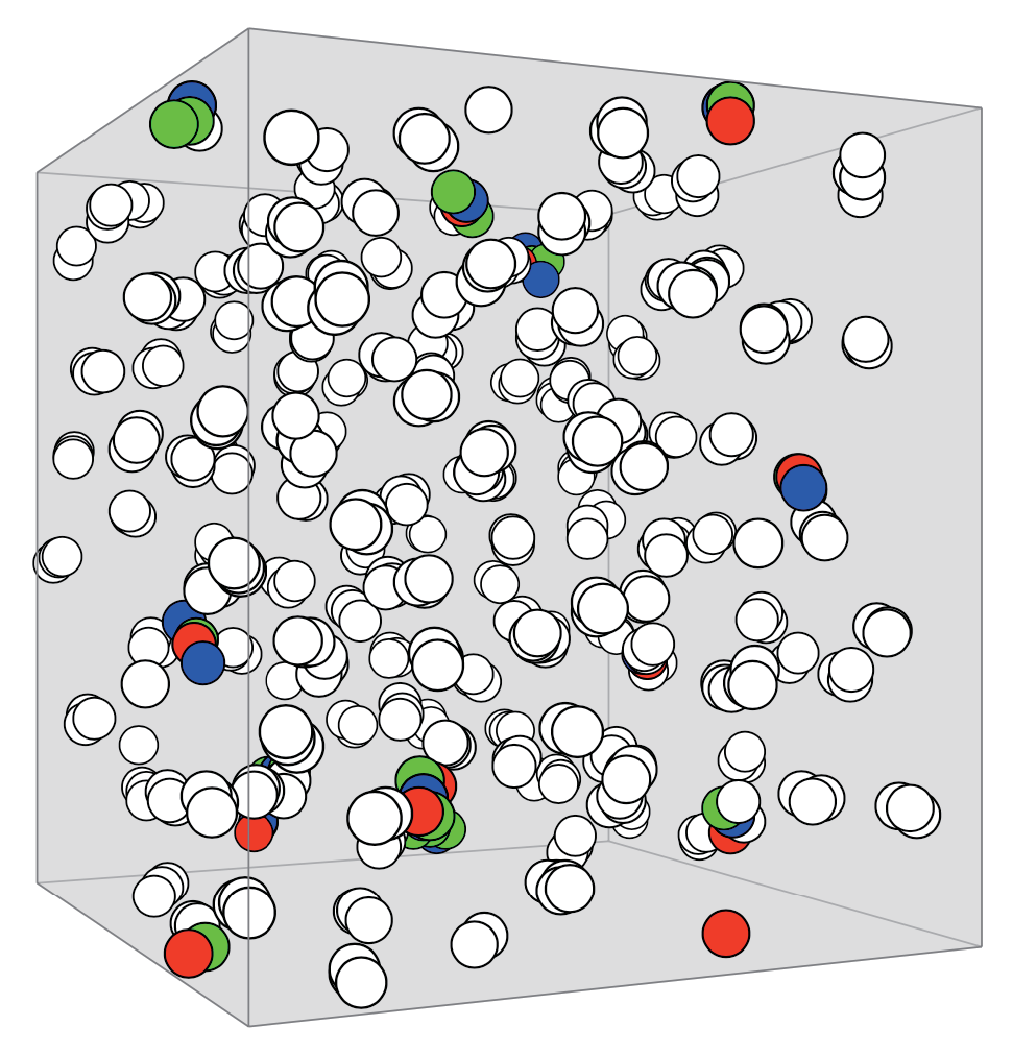}
\includegraphics[width=0.22\textwidth]{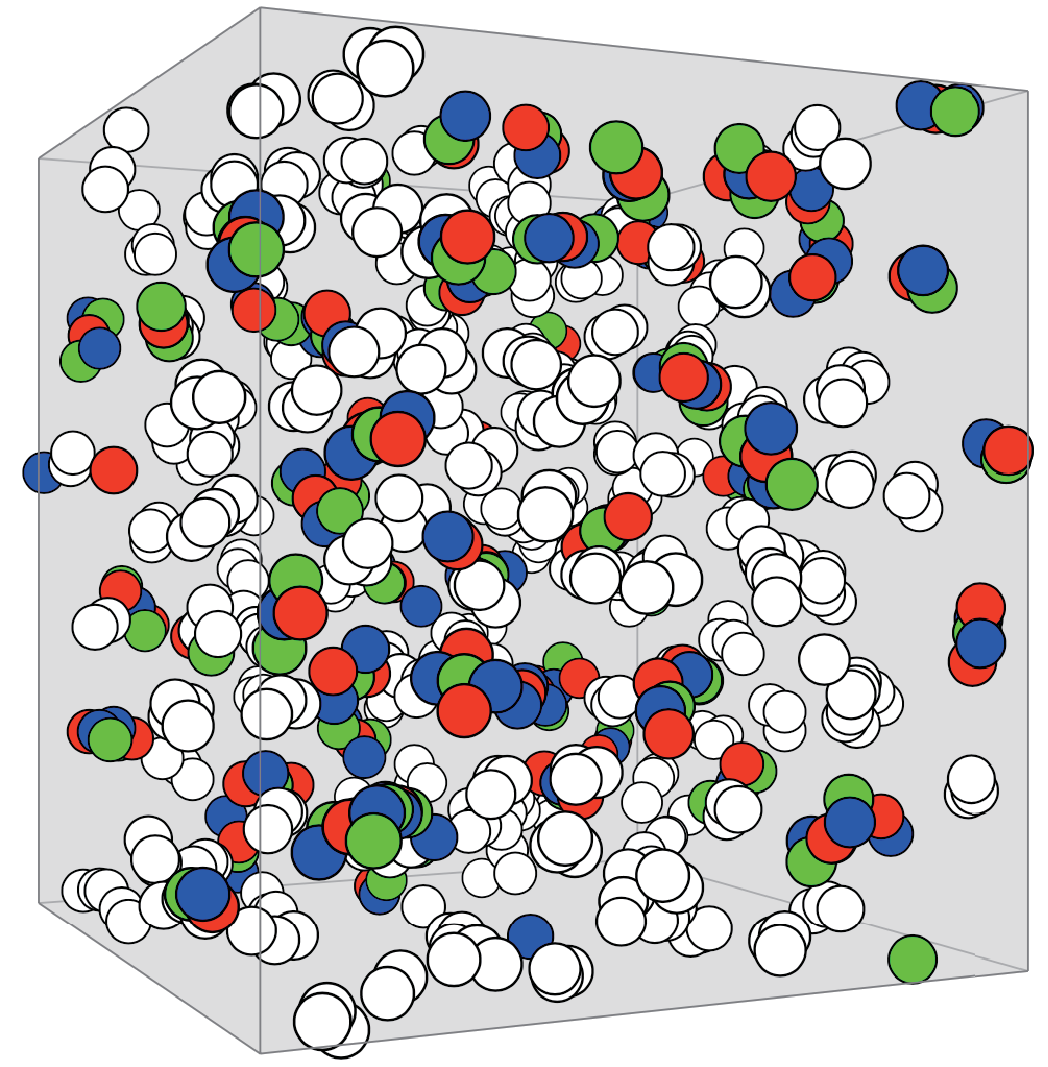}
\includegraphics[width=0.22\textwidth]{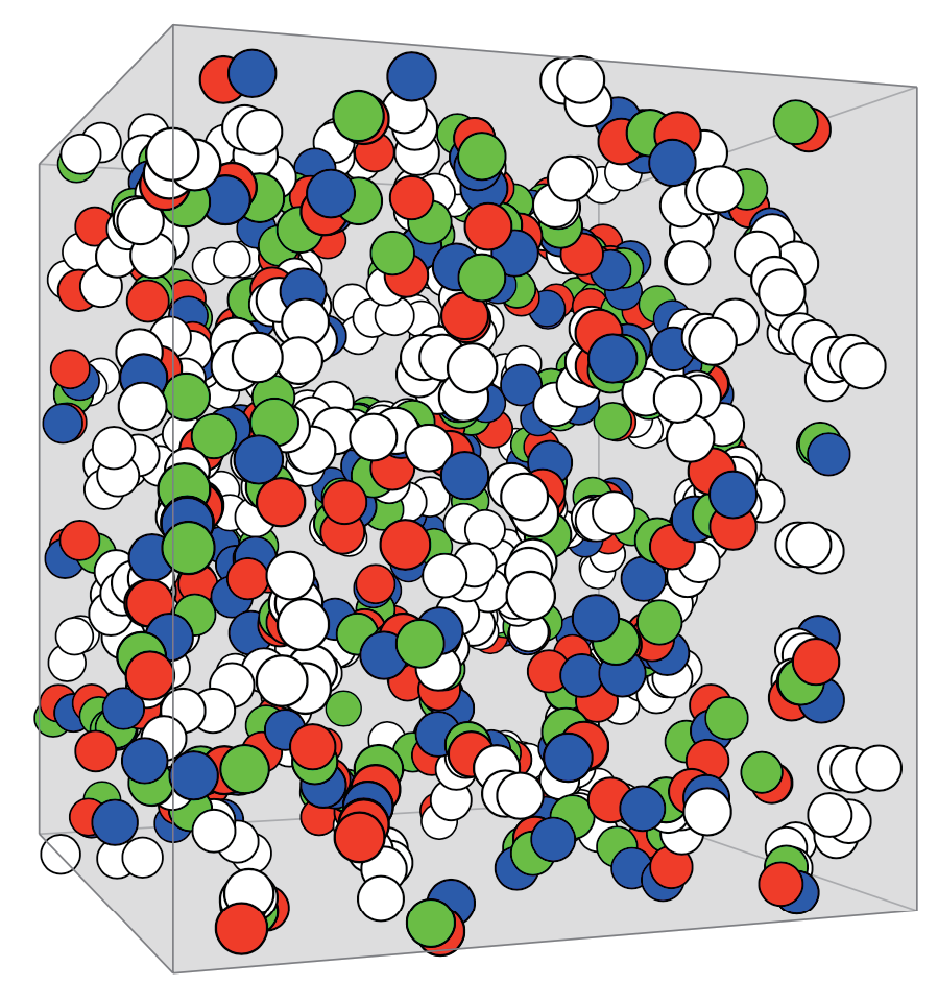}
\includegraphics[width=0.23\textwidth]{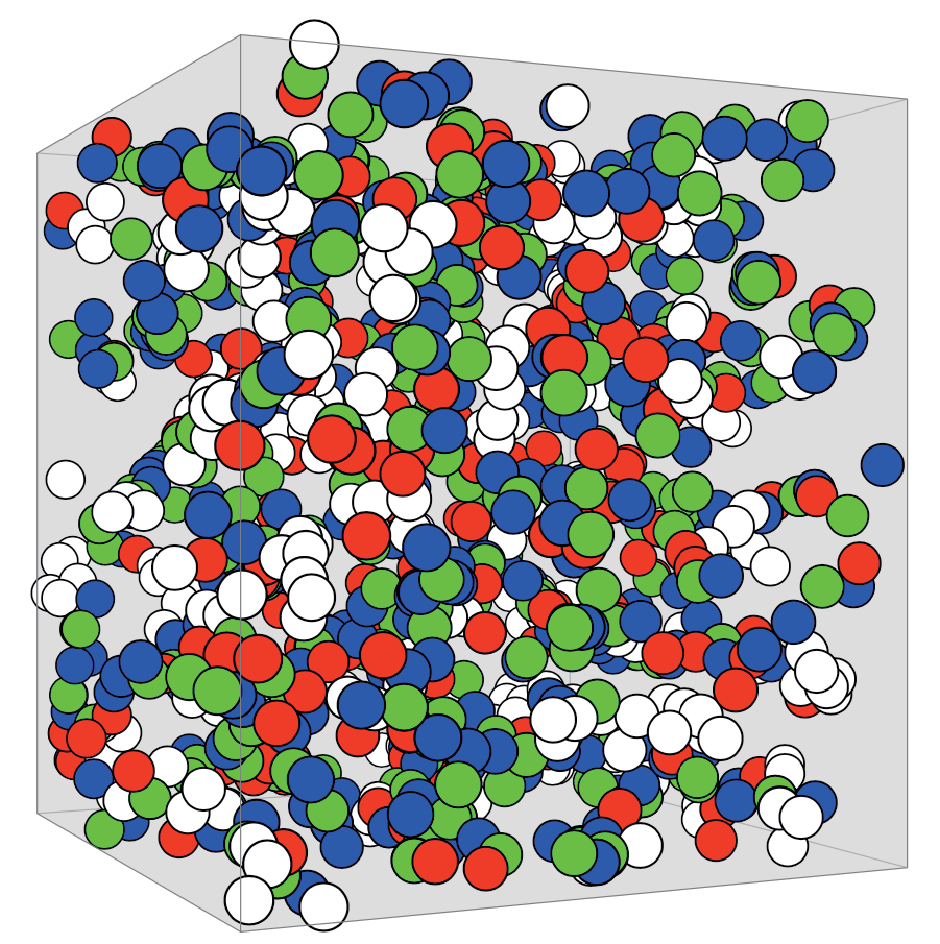}
\par
\vspace{-2.5mm}
\hspace{27mm} 0.833 fm$^{-3}$ \hspace{27mm} 1.166 fm$^{-3}$\hspace{24mm} 1.583 fm$^{-3}$ \hspace{24mm} 1.916 fm$^{-3}$
\caption{\label{fig:udd_cmd} Same as Fig.~\ref{fig:ud_cmd}, but for $udd$ matter.}
\end{figure*}

The conditions for confinement/deconfinement are set as follows: If three quarks are within a certain distance $d_{\rm cluster}$
 and the total color of three quarks is white with an accuracy $\varepsilon$, these quarks are considered confined.
It is formulated as
\begin{eqnarray}
\left\{
\begin{array}{l}
\left |{\bf R}_i-{\bf R}_j\right |<d_{\rm cluster}  ~~~~~(i,j=1,2,3),\\
\displaystyle{ \sum_{a=1}^8\left[\sum_{i=1}^3 \langle\chi_i|\lambda^a|\chi_i\rangle\right]^2 <\varepsilon }.
\end{array}
\right.
\end{eqnarray}
The values of the criteria are set as $d_{\rm cluster} = 0.33 $~fm and $\varepsilon = 0.01$ in this study.
The perspectives of $ud$ matter and $udd$ matter depended on the density are shown in Fig.\ref{fig:ud_cmd} and Fig.\ref{fig:udd_cmd}. 
In these figures, the colors of the quarks are represented using the discontinuous colors red, blue, and green~(RGB) for visibility, however in the actual calculations, the quarks are in a mixed state of RGB reflecting the internal degrees of freedom in Eq. (\ref{eq:color}). 
Namely, the color is selected and depicted among RGB, which is closest to the color mixed state for each particle.
Particles with white colors indicate quarks in the confined state, while red, blue, and green ones show in the deconfined quark matter. 
In these figures, it can be seen that the fraction of deconfined quarks gradually increases with density.
This behavior is similar to the percolative depiction~\cite{Kojo2015}.

\begin{figure}[tb]
\includegraphics[width=0.47\textwidth]{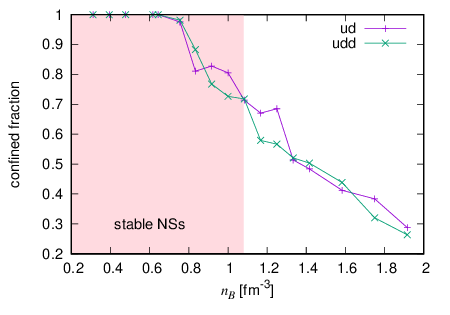}
\caption{\label{fig: fraction} Fractions of quark number within baryon state against total quark number for $ud$ matter and $udd$ matter, shown in  Fig.~\ref{fig:ud_cmd} and Fig.~\ref{fig:udd_cmd}.
The painted area indicates the region below the maximum density 1.08 fm$^{-3}$, i.e., at the density allowed in the interior of stable NSs.}
\end{figure}

The fractions of confined quarks to total quarks in $ud$ and $udd$ matter are shown in Fig.\ref{fig: fraction}. 
For both matters, deconfined quarks appear around 0.6 fm$^{-3}$. 
As we will show later, the maximum density in stable neutron stars is 1.08 fm$^{-3}$, hence our calculations suggest that the interior of neutron stars is mostly filled with confined quark matter, i.e., baryons. The fraction of deconfined quark matter is roughly estimated to be less than 0.3 in neutron stars.

\subsection{MASS-RADIUS RELATION}

\begin{figure}[tb]
\includegraphics[width=0.49\textwidth]{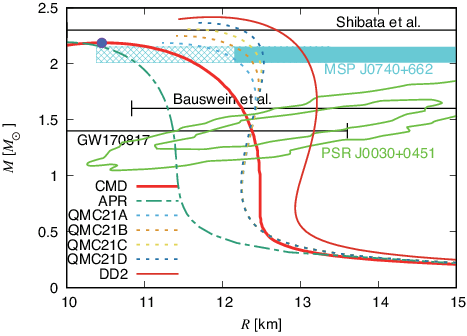}
\caption{\label{fig: MR} 
The mass-radius relation obtained from the numerical results of CMD, where the maximum mass and the central baryon number density of NSs are $M_{\rm max}= 2.19 M_\odot$ and $n_B=$ 1.08 fm$^{-3}$, represented by the filled color circle. For comparison, the results by Akmal {\it et al.}(APR)~\cite{APR}, Kojo {\it et al.}(QHC21A-D)~\cite{QHC21A}, and S. Typel {\it et al.}(DD2)~\cite{DD2} are also included.
The constraints obtained from the astronomical observations with NICER, PSR J0030+0451~\cite{Riley2019, Miller2019}, and MSP J0740+6620~\cite{Riley2021, Miller2021}, are also shown. The meshed area on the constraint of MSP J0740+6620 has been obtained from the NICER-only analysis, and the painted area was given based on NICER+XMM-Newton datasets. We also show the constraints inferred from event GW170817, labeled by ``GW170817", ``Bauswein {\it et al.}", and ``Shibata {\it et al.}". 
As for ``Bauswein {\it et al.}" , they give a lower limit of NS radius as $R_{1.6M_\odot}$=10.3-10.7 km with the mass of $M=1.6 M_\odot$~\cite{Bauswein2017}. 
As for ``Shibata {\it et al.}", they give an upper limit on the NS mass of $M<2.3 M_\odot$ by comparing their numerical relativistic simulations with the electromagnetic (EM) observation accompanied by GW170817~\cite{Shibata2019}.
 See the text and the reference for the details. }
\end{figure}

Finally, the mass-radius (MR) relation for neutron stars obtained by CMD calculation is shown Fig.~\ref{fig: MR}. 
The results by Akmal {\it et al.}(APR)~\cite{APR}, Kojo {\it et al.}(QHC21A-D)~\cite{QHC21A}, and S. Typel {\it et al.}(DD2)~\cite{DD2} are also included for comparison.
The radii obtained from CMD calculations are larger than those from the two models, but still consistent with the constraints imposed by astronomical observations, as shown below.

The constraints obtained from the astronomical observations with NICER, PSR J0030+0451~\cite{Miller2019}, and MSP J0740+6620~\cite{Riley2021, Miller2021}, are also shown in the figure. The area bounded by two jagged circles represents 1$\sigma$ and 2$\sigma$ ranges of the observational result based on the analysis assuming three uniform-temperature oval spots~\cite{Miller2019}. Note that these ranges agree with the other result assuming two uniform-temperature oval spots. As for PSR J0030+0451, see also Ref.~\cite{Riley2019}.
The meshed area on the constraint of MSP J0740+6620, $R={11.5}_{-1.2}^{+1.8}$ km~(68\%), has been obtained from the NICER-only analysis, and the painted area, $R={13.7}_{-1.5}^{+2.6}$ km~(68\%), were given based on NICER+XMM-Newton datasets~\cite{Miller2021}. The mass of MSP J0740+6620 has been suggested as ${2.08 \pm 0.07} M_\odot$. Our CMD calculations give the maximum mass 2.19 $M_\odot$, which is over the lower limit of the mass suggested from MSP J0740+6620. The maximum density in stable neutron stars is $n_B$ = 1.08 fm$^{-3}$ in our CMD calculations, which indicates that the interior of neutron stars is occupied by mostly hadrons, as shown in Fig.~\ref{fig: fraction}.
Note that our result lies outside the NICER+XMM-Newton range (painted area). We believe that it would be satisfied after taking into account the other physics, which should be improved in the future, such as color-magnetic interactions.

Besides the above two constraints from NICER, we also take into account the observations related to event GW170817, and show them in Fig.~\ref{fig: MR}. 

First, the gravitational wave observation, GW170817, itself gave us valuable information on the tidal deformability of the neutron stars, which could be a constraint for the radii~\cite{Abbott2017}. The observation suggests that the radii should be less than 13.6 km for neutron stars with a canonical mass, 1.4 $M_\odot$. The dimensionless tidal deformability from our CMD calculation is $ \Lambda_{1.4 M\odot}=458$, and the corresponding radius is less than the constraint by GW170817. 

The electromagnetic (EM) observation accompanied by GW170817 has also provided complementary information: the merger was suggested not to be in a prompt collapse to a black hole (BH) because of the large quantity of ejecta and its high electron fraction. 
Given that the threshold for prompt collapse depended on the NS compactness, Bauswein {\it et al.} had placed a lower limit of NS radius as  $R_{1.6M_\odot}$=10.3-10.7 km with the mass of $M=1.6 M_\odot$~\cite{Bauswein2017}. Moreover, comparing numerical relativistic simulations as for a supermassive neutron star remnant with the measured binary mass $M_{\rm tot} = 2.74+0.04 M_\odot$~\cite{Abbott2017}, 
Shibata {\it et al.} have given an upper limit on the NS mass of $M<2.3 M_\odot$~\cite{Shibata2019}. 
For more details on EM counterparts, see also the review by Metzger~\cite{Metzger2020} and the reference therein. 

Although not included in Fig.~\ref{fig: MR} to avoid complications, it should be mentioned that Margalit and Metzger gave a more strict constraint, $ M<2.17 M_\odot$~\cite{MargalitMetzger2017}. On the other hand, Keck-telescope spectrophotometry and imaging of the companion of the ``black widow" pulsar PSR J09520607, suggests 2.35 $\pm 0.17 M_\odot$~\cite{Romani2022}. Therefore, careful consideration is still needed regarding the maximum mass of neutron stars. With this background in mind, this paper adopts the more conservative constraint by Shibata {\it et al.}~\cite{Shibata2019} rather than the one by Margalit and Metzger~\cite{MargalitMetzger2017}.

\section{DISCUSSION}
In this study, CMD calculations were performed to obtain the EOS, which is consistent with the constraints from the astrophysical observations. At low densities, our results deviate slightly from the values suggested by various nuclear experiments: our model satisfies just barely the range of $K$ by Danielewicz {\it et al.}\cite{Danielewicz2002} but not for the others~\cite{Piekarewicz2004, Colo2004, Sturm2001, Hartnack2006}, while it does the fiducial values of $n_0$, $J$, and $L$~\cite{Tsang2012,Lattimer2013}.

In our current framework, the result suggests that the deconfinement of quarks is crossover in stable neutron stars. 
Moreover, the insides of NSs are occupied mainly by confined hadron matter. This is because the EOS inevitably needs to become stiff to support neutron stars with masses exceeding two solar mass to be consistent with the astronomical observations. As a consequence, the core of the neutron star does not achieve a sufficiently high density. In any case, until the sign problem is resolved, there is no means to obtain information about high-density matter other than by assuming  some frameworks and obtaining insights from astronomical observations. Within the limited framework of our current model, it implies that the inclusion of quark many-body effects is necessary at least: it is inevitable to introduce $\epsilon_\omega$, $\epsilon_\sigma$ for reproducing observational results in this work. The similar discussions are also found in the other EOS models, where the nonlinear many-body effects are required to be consistent with the observations~\cite{DD2, Muto2021, Yamamoto2022}.

Note that this result is subject to further refinement in our ongoing research, since we still miss a lot of physics: we adopt simple Newtonian interactions without strangeness, quarks spin-spin interactions (color-magnetic interactions), etc. 
Therefore, changing our theoretical framework would certainly alter the value of maximum density, 1.08 fm$^{-3}$. For instance, 
one can easily anticipate that the EOS would be softened when we further take into account strangeness, but it is unclear how relativistic interactions and color magnetic interactions would change the EOS on the other side. Particularly, the relativistic effects are expected to have a significant impact on the speed of sound. In this study, the speed of sound has not  violated the causality for densities up to 1.08 fm$^{-3}$, but for near the density, 1.42 fm$^{-3}$. Hence, our current conclusion is not crucial. In the future, we anticipate that properly considering relativistic effects will lead to more realistic results on the speed of sound. 

Note that our model is based on the constituent quark models, and then we do not take into account the chiral condensations. It is a future challenge to incorporate such physics into our CMD code. Furthermore, in this study, instead of considering antisymmetrization, we simply introduced effective coupling constants $\kappa_{\rm eff}$ and $\alpha_S^{\rm eff}$ \cite{MaruyamaHatsuda2000, Akimura2005}. However, it should be carried out to compare our method with AMD for the evaluation of our approach~\cite{Enyo2005,Enyo2007}. Nevertheless, antisymmetrization itself incurs a computational cost of $N^4$, so it would be more practical to perform calculations in finite systems for such comparison. It means that we also need to consider the antiparticles, neglected in the present study since we focus on EOSs at zero temperature. Moreover, since gluons themselves also play important roles at quark-gluon-plasma in heavy-ion collisions, they should be taken into account. 

Thus, our model is in the early stages of development, and we aim to address the aforementioned issues by comparing it with various other models, not only with AMD. Indeed, various approaches have been pursued in theoretical simulations of heavy-ion collisions, as mentioned above. In the state of the art, the initial nonequilibrium dynamics of the collision are treated by particle-based models such as UrQMD and JAM, and then it is switched to hydrodynamics under thermalized conditions\cite{Steinheimer2008, Petersen2008, Akamatsu2018}. 
It would be intriguing to compare our model with successful models such as those mentioned above, and we strongly look forward to conducting such comparisons in the future.

\begin{acknowledgments}
We are grateful to N.~Hoshi, Y.~Mukobara, H.~Tanihisa, S.~Ozaki, C. J. Xia, P.~Gubler, K.~Kyutoku, T. Muto, A. Park, S-H. Lee, M.~Oka, and T.~Hatsuda for fruitful discussions.
This work was supported by JSPS KAKENHI Grants No., 20H04742, No. 20K03951.
The calculations were performed by the supercomputing system HPE SGI 8600 at the Japan Atomic Energy Agency.
\end{acknowledgments}

\newpage

\appendix

\section{BOX SIZE DEPENDENCE}

\begin{figure}[tb]
\includegraphics[width=0.52\textwidth]{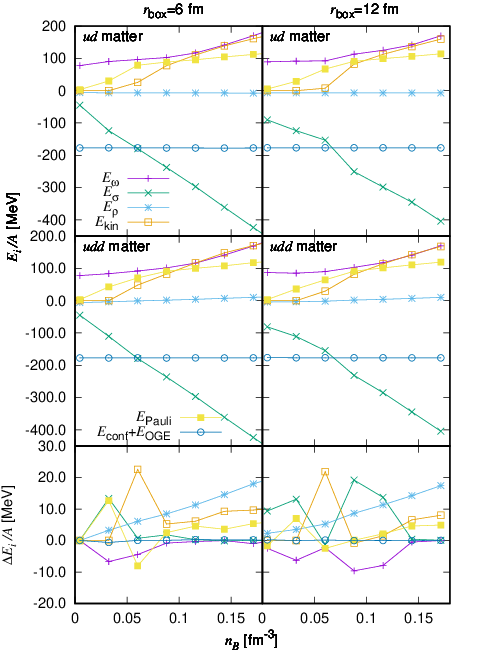}
\caption{\label{fig:boxsize} Comparison of energy components for the cases where $r_{\rm{box}}$ is 6~fm (left panels) and 12~fm(right panels). The meaning of each line type is the same as in Fig.\ref{fig: Ei}. Additionally, the upper and middle panels represent $ud$ and $udd$  matter, and the lower panels illustrate the difference between the both matters. } 
\end{figure}

In this appendix, we discuss the size dependence of periodic boundary conditions for CMD calculations. In Fig.~\ref{fig:boxsize}, the energy components for the box size $r_{\rm{box}}$ = 6 fm, employed in this study, are presented, along with the results for a box size twice as large for comparison. 

To explain the box size dependence, let us first focus on the vicinity of zero density. Considering the case of a single baryon ($N$=3) with $r_{\rm{box}}$=6~fm, due to periodic boundary conditions, the interbaryon distance is limited to 6~fm. In contrast, for the case of $r_{\rm{box}}$=12~fm with the same density, there are eight baryons in the box. Through variational calculations, the inter-particle distances are optimized, and as a result, 
they can be different from 6~fm though the average is 6~fm.
Therefore, when we compare these results, there are the differences in the energy components due to the variations in interbaryon distances. This box size dependence arises for the other densities as well, based on the mean interbaryon distance differences.

Due to this dependence, low-density data in this study is considered less reliable and has been omitted. 
On the other hand, as the density increases, the average interbaryon distances become similar regardless of the box size. 
Consequently, each energy contribution tends to show similar values. 
Additionally, as mentioned earlier, for the case of $r_{\rm{box}}$=6~fm with $N$=3, the interbaryon distance is equal to $r_{\rm{box}}$. However, this value is large enough that the $E/A$ for $N$=3 in the box shows the same value with the calculation result for the finite case (without the boundary), namely 938 MeV. Therefore, in accordance with convention, the values of $E/A$ in Fig.\ref{fig: eos} are referenced to a new baseline, deducting the value 938 MeV.

%

\newpage



\nocite{*}

\begin{thebibliography}{99}

\bibitem{Antoniadis2013}
J. Antoniadis {\it et al.}, Science {\bf 340}, 1233232 (2013).

\bibitem{Abbott2017}
B. P. Abbott {\it et al.} (The LIGO Scientific Collaboration and the Virgo Collaboration), Phys. Rev. Lett. {\bf 119}, 161101 (2017).

\bibitem{Bauswein2017}
A. Bauswein, O. Just, H-T. Janka, and N. Stergioulas, Astrophys.\ J.\ Lett. {\bf 850}, L34 (2017).

\bibitem{Riley2019}
T. E. Riley {\it et al.}, Astrophys. J. {\bf 887}, L21, (2019).
\bibitem{Miller2019}
M. C. Miller {\it et al.}, Astrophys. J. {\bf 887}, L24, (2019).
\bibitem{Riley2021}
T. E. Riley {\it et al.}, Astrophys. J. {\bf 918}, L27, (2021).
\bibitem{Miller2021}
M. C. Miller {\it et al.}, Astrophys. J. {\bf 918}, L28, (2021).


\bibitem{Yasutake2009}
N. Yasutake, T. Maruyama, and T. Tatsumi, Phys. Rev. D{\bf 80}, 123009 (2009).

\bibitem{Chen2013}
H. Chen, G. F. Burgio, H.-J. Schulze, and N. Yasutake, Astron. \& Astrophys. {\bf 551}, A13 (2013).

\bibitem{Yasutake2014}
N. Yasutake {\it et al.}, Phys. Rev. C{\bf 89}, 065803 (2014).

\bibitem{Maslov2019}
K. Maslov {\it et al.}, Phys. Rev. C{\bf 100}, 025802 (2019).

\bibitem{Xia2019}
C.-J. Xia, T. Maruyama, N. Yasutake, and T. Tatsumi, Phys. Rev. D{\bf 99}, 103017 (2019).

\bibitem{Xia2020}
C.-J. Xia, T. Maruyama, N. Yasutake, T. Tatsumi, H. Shen, and H. Togashi, Phys. Rev. D{\bf 102}, 023031 (2020).

\bibitem{Masuda2016}
K. Masuda, T. Hatsuda, and T. Takatsuka, Eur. Phys. J. A {\bf 52}, 65 (2016).

\bibitem{Baym2019}
G. Baym, S. Furusawa, T. Hatsuda, T. Kojo, and H. Togashi, Astrophys. J. {\bf 885}, 42 (2019).

\bibitem{Minamikawa2021}
T. Minamikawa, T. Kojo, and M. Harada, Phys. Rev. C {\bf 103}, 045205 (2021).

\bibitem{Blaschke2022}
D. Blaschke, E. -O. Hanu, and S. Liebing, Phys. Rev. C {\bf 105}, 035804 (2022).

\bibitem{QHC21A}
T. Kojo, G. Baym, and T. Hatsuda, Astrophys. J. {\bf 934}, 46 (2022).

\bibitem{Tsang2012}
M. B. Tsang {\it et al.}, Phys. Rev. C {\bf 86}, 015803 (2012).

\bibitem{Lattimer2013}
J. M. Lattimer and Y. Lim, Astrophys. J. {\bf 771}, 51 (2013).

\bibitem{Vinas2014}
X. Vi\~{n}as, M. Centelles, X. Roca-Maza, and M. Warda, Eur. Phys. J. A {\bf 50}, 27 (2014).


\bibitem{Li2019}
B.-A. Li, P. G. Krastev, D.-H. Wen, and N.-B. Zhang, Eur. Phys. J. A {\bf 55}, 117 (2019). 

\bibitem{Danielewicz2017}
P. Danielewicz, P. Singh and J. Lee, Nucl. Phys. A {\bf 958}, 147 (2017).  

\bibitem{Estee2021}
J. Estee {\it et al.}, Phys. Rev. Lett. {\bf 126}, 162701 (2021). 

\bibitem{Reed2021}
B. T. Reed, F. J. Fattoyev, C. J. Horowitz, and J. Piekarewicz, Phys. Rev. Lett. {\bf 126}, 172503 (2021) 

\bibitem{DD2}
S. Typel, G. R\"{o}pke, T. Kl\"{a}hn, D. Blaschke, and H. H. Wolter, Phys. Rev. C {\bf 81}, 015803 (2010) 

\bibitem{Tanaka2021}
J. Tanaka {\it et al.}, Science {\bf 371}, 260 (2021) 

\bibitem{Danielewicz2002}
P. Danielewicz, R. Lacey, and W. G. Lynch, Science, {\bf 298}, 1592 (2002). 

\bibitem{Piekarewicz2004}
J. Piekarewicz, Phys. Rev. C {\bf 69}, 041301 (2004). 

\bibitem{Colo2004}
G. Col\`{o} {\it et al.}, Phys. Rev. C {\bf 70}, 024307 (2004). 

\bibitem{Sturm2001}
C. Sturm {\it et al.}, Phys. Rev. Lett. {\bf 86}, 39 (2001). 

\bibitem{Hartnack2006} 
Ch. Hartnack {\it et al.}, Phys. Rev. Lett. {\bf 96}, 012302 (2006). 

\bibitem{Lonardoni2020}
D. Lonardoni, I. Tews, I., S. Gandolfi, and J. Carlson, Phys. Rev. Res. 2, 022033 (2020).

\bibitem{Drischler2021a}
C. Drischler, S. Han, J. M. Lattimer {\it et al.}, Phys. Rev. C {\bf 103}, 045808 (2021).

\bibitem{Drischler2021b}
C. Drischler, J. W. Holt, and C. Wellenhofer, Annu. Rev. Nucl. Part. Sci. 71, 403 (2021).

\bibitem{MaruyamaHatsuda2000} 
T. Maruyama and T. Hatsuda, Phys. Rev. {\bf C 61}, 062201 (2000).

\bibitem{Akimura2005} 
Y. Akimura, T. Maruyama, N. Yoshinaga, and S. Chiba, Eur. Phys. J. A {\bf 25}, 405 (2005).


\bibitem{Yoshimoto2000}
T. Yoshimoto, T. Sato, M. Arima, and T.S.H. Lee, Phys. Rev. C {\bf 61}, 065203 (2000).

\bibitem{Feldmeier1990}
H. Feldmeier, Nucl. Phys. {\bf A515}, 147 (1990).

\bibitem{Ono2005}
A. Ono, H. Horiuchi, T. Maruyama, and A. Ohnishi, Prog. Theor. Phys. {\bf 87}, 1185 (1992).

\bibitem{Guichon1988}
P. A. M. Guichon, Phys. Lett. B {\bf 200}, 235 (1988).

\bibitem{Caplan2018}
M. E. Caplan, A. S. Schneider, C. J. Horowitz, Phys. Rev. Lett., {\bf 121}, 132701 (2018).


\bibitem{Cheng1999}
H. Cheng, L. Greengard, and V. Rokhlin, J. Comput. Phys. {\bf 155} 468 (1999).


\bibitem{Baym1971}
G. Baym, C. Pethick, and P. Sutherland, Astrophys. J. {\bf 170}, 299 (1971).

\bibitem{APR}
A. Akmal, V. R. Pandharipande and D. G. Ravenhall, Phys. Rev. C {\bf 58} 1804 (1998).


\bibitem{Comp}
https://compose.obspm.fr/home/

\bibitem{Legred2021}
I. Legred, K. Chatziioannou, R. Essick, S. Han, and P. Landry, Phys. Rev. D {\bf 104}, 063003 (2021).






\bibitem{Kojo2015}
T. Kojo, P. D. Powell, Y. Song, and G. Baym, Phys. Rev. D {\bf 91}, 045003 (2015).


\bibitem{Shibata2019}
M. Shibata, E. Zhou, K. Kiuchi, and S. Fujibayashi, Phys. Rev. D {\bf 100}, 023015 (2019).
\bibitem{Metzger2020}
B. D. Metzger, Living Rev. Relativity, 23, 1 (2020). 

\bibitem{MargalitMetzger2017}
B. Margalit and B. D. Metzger, Astrophys. J. Lett. {\bf 850}, L19 (2017).

\bibitem{Romani2022}
R. W. Romani, D. Kandel, A. V. Filippenko, T. G. Brink, and W. Zheng, Astrophys. J. Lett. {\bf 934}, L17 (2022).

\bibitem{Muto2021}
T. Muto, T. Maruyama, and T. Tatsumi, Phys. Lett. B {\bf 820}, 136587 (2021).
\bibitem{Yamamoto2022}
Y. Yamamoto, N. Yasutake, and Th. A. Rijken, Phys. Rev. C {\bf 105}, 015804 (2022).

\bibitem{Enyo2005}
Y. Kanada-En'yo, O. Morimatsu, and T. Nishikawa, Phys.  Rev. C {\bf 71}, 045202 (2005).
\bibitem{Enyo2007}
Y. Kanada-En'yo, O. Morimatsu, and T. Nishikawa, Prog. Theor. Phys. Suppl. 168, 194 (2007).

\bibitem{Steinheimer2008}
J. Steinheimer, M. Bleicher, H. Petersen, S. Schramm,
H. Stocker, and D. Zschiesche, Phys. Rev. C {\bf 77}, 034901 (2008). 
\bibitem{Petersen2008}
H. Petersen, J. Steinheimer, G. Burau, M. Bleicher, and
H. Stocker, Phys. Rev. C {\bf 78}, 044901 (2008).
\bibitem{Akamatsu2018}
Y. Akamatsu, M. Asakawa, T. Hirano, M. Kitazawa, K. Morita, K. Murase, Y. Nara, C. Nonaka, and A. Ohnishi, Phys. Rev. C {\bf 98}, 024909 (2018).


\end{thebibliography}
\end{document}